\documentstyle[aas2pp4]{article}
\newcommand{\be}{\begin{equation}}
\newcommand{\ba}{\begin{eqnarray}}
\newcommand{\ee}{\end{equation}}
\newcommand{\ea}{\end{eqnarray}}
\newcommand{\sol}{M_{\sun}}

\newcommand{\etal}{{\it et al. }}

\slugcomment{Kip's 5 April 1997 Revision}

\lefthead{Nakamura et al. }
\righthead{}

\begin{document}

\title{Gravitational Waves from Coalescing Black Hole MACHO Binaries}

\author{Takashi Nakamura}
\affil{Yukawa Institute for Theoretical Physics, Kyoto University,
Kyoto 606}
\vspace{5mm}

\author{Misao Sasaki  and Takahiro Tanaka }
\affil{Department of Earth and Space Science, Osaka  University,
Toyonaka 560}
\and
\author{Kip S. Thorne }
\affil{Theoretical Astrophysics,
California Institute of Technology, Pasadena, California 91125}

\received{}
\accepted{}

\begin{abstract}
 If  MACHOs are black holes of mass  $\sim0.5\sol$, they must have been
formed in the early universe when the temperature was $\sim$1 GeV.
We estimate that in this case in our galaxy's halo out to $\sim$ 50kpc there 
exist $\sim 5\times 10^{8}$ black hole binaries whose
coalescence times are comparable to the age of the universe, so that the
coalescence rate will be $\sim 5\times 10 ^{-2}$ events/year/galaxy. 
This suggests that we
can expect a few events/year within 15Mpc. The gravitational waves
from such coalescing black hole MACHOs can be detected by the
first generation of interferometers in the LIGO/VIRGO/TAMA/GEO network. 
Therefore, the existence of black hole MACHOs can be tested within the next
five years by gravitational waves.    
( submitted to Apj Letters April 11 1997) 
\end{abstract}


\section{Introduction}
  The analysis of the first 2.1 years of photometry 
of 8.5 million stars in the Large Magellanic 
Cloud (LMC) by the MACHO collaboration (\cite{alco96a})
 suggests that $0.62^{+0.3}_{-0.2}$ of the halo consists of MACHOs of
 mass $0.5^{+0.3}_{-0.2}\sol$ 
 in the standard spherical flat rotation halo 
model. The preliminary analysis of four years of data suggests the
existence of at least four additional microlensing events with
$t_{dur}\sim 90$days in the direction of the LMC (\cite{prat97}).
The estimated masses of these MACHOs are just the mass of 
 red dwarfs. However, the contribution of the  halo red dwarfs to MACHO 
events should be small  since  the observed density of halo red dwarfs 
is too low (\cite{bahc94,graf96a,graf96b}).
As for white dwarf MACHOs,  
 the mass fraction of white dwarfs in the halo should be less than 10\%
since, assuming the Salpeter initial mass function (IMF), the bright 
progenitors of more
white dwarfs than this would be in conflict with the number counts of 
distant galaxies
(\cite{char95}). If the IMF has a sharp peak around
$2\sol$, then the fraction could be 50\% or so (\cite{adam96}), sufficient to
explain the MACHO observations. 
The existence
of such a population of halo white dwarfs may or may not be
consistent with the  observed luminosity function
(\cite{goul97,lidm97,free97}). In any case, 
future observations of high velocity white dwarfs in our solar
neighborhood (\cite{lidm97}) will prove whether white dwarf
MACHOs can exist or not. 

If the  number of high velocity white dwarfs turns out to be large
enough to explain the MACHOs, then stellar formation theory should explain 
why the IMF is sharply peaked at   
$\sim 2\sol$. If it is not, there arises a real possibility that
MACHOs are absolutely new objects such as black holes of mass 
 $\sim$ 0.5$\sol$ which could only be
formed in the early universe, or boson stars with the mass of the
boson $\sim 10^{-10}$eV. Of course it is still possible that
an overdense clump of MACHOs exists toward the LMC (\cite{naka96}), MACHOs are
brown dwarfs in the rotating halo (\cite{spir97}) or MACHOs are stars 
in the thick disk (\cite{turn97}). 

In this {\it Letter} we consider the case of black hole
MACHOs (BHMACHOs). In this case, there must be a huge number 
(at least $\sim 4\times 10^{11}$) of black holes in the halo and it is 
natural to expect that some of them are binaries. 
In \S2 we estimate the fraction $f(a,e)dade$ of all BHMACHOs that are
in binaries with semimajor axis $a$ in range $da$ and eccentricity $e$ in $de$.
We then use this distribution to estimate two observable event rates:
First (end of \S2) the rate of microlensing events we should expect toward
the LMC due to binaries with separation $\gtrsim 2\times
10^{14}$cm; our result is in accord with the observation of one such event thus
far (\cite{Benn96}).  Second (\S3), the rate of coalescence of BHMACHO binaries 
out to 15Mpc distance.  The gravitational waves from such coalescences should
be detectable by the first interfereomters in the LIGO/VIRGO/TAMA/GEO 
network (\cite{ligo,virgo,tama,geo}), and our estimated event rate is a few
events per year.  In \S 4 we discuss some implications of our estimates.

\section{Formation of Solar Mass Black Hole MACHO Binaries}
Since it is impossible to make a black hole of mass $\sim 0.5\sol$
as a product of stellar evolution, we must consider the
formation of solar mass black holes in the very early universe 
(\cite{yoko95,jeda97}).  Our viewpoint here, however, 
is not to study detailed formation mechanisms, but to estimate the binary
distribution that results.  

The density parameter of BHMACHOs, $\Omega_{BHM}$, must 
be comparable to $\Omega_{b}$ (or $\Omega_{CDM}$) 
to explain the number of observed MACHO events.  
For simplicity, we assume that BHMACHOs 
dominate the matter 
energy density, i.e., $\Omega=\Omega_{BHM}$, 
though it is possible to consider other dark
matter components in addition to BHMACHOs.
To determine the mean separation of the BHMACHOs, 
it is convenient to consider it at the time of  
matter-radiation equality, $t=t_{eq}$. 
At this time, the energy densities of radiation and BHMACHOs are approximately
equal and are given by 
$
 \rho_{eq}=1.4\times 10^{-15} \left(\Omega h^2\right)^4
\hbox{\rm g/cm$^3$},
$
where $h$ is the Hubble parameter in units of 100km/s/Mpc. 
Correspondingly, the mean separation of BHMACHOs with mass $M_{BH}$ 
at this time is given by 
\ba
 \bar x & = & \left( {M_{BH}/ \rho_{eq}}\right)^{1\over 3}\cr
        & = & 1.1\times 10^{16}\left({M_{BH}/\sol}\right)^{1\over 3}
   (\Omega h^2)^{-{4\over 3}}\hbox{\rm cm}.
\ea
We set the scale factor $R$ to unity at $t=t_{eq}$, 
so $\bar x$ can also be regarded as the comoving mean separation. 
Note that the Hubble horizon scale at $t=t_{eq}$ is
$
L_{eq} \sim \sqrt{3 c^2/8\pi G \rho_{eq}}
   = 1.1 \times 10^{21}\left(\Omega h^2\right)^{-2} \hbox{\rm cm}.
$

During the radiation dominated era, the total energy inside 
the horizon increases as $R^2$. 
Since the Jeans mass in this era is essentially the 
same as the horizon mass, black holes 
are formed only at  
the time when the horizon scale is equal to the Schwarzschild radius 
of a BHMACHO. Thus the scale factor at the 
formation epoch becomes 
\be
 R_f=\sqrt{{G M_{BH}\over c^2 L_{eq}}}
 =1.2\times 10^{-8}\left({M_{BH}\over \sol}\right)^{1\over 2}
 \left(\Omega h^2\right).   
\label{Rf}
\ee
 The age of the universe and the temperature $T_f$ at $R=R_f$
are $\sim 10^{-5}$ sec and $\sim 1$ Gev, respectively.

As a foundation for computing the distribution function $f(a,e)$ for BHMACHO
binaries, we assume that the BHMACHOs are created with a
distribution of comoving separations $x$ that is uniform over the range from an
initial physical separation equal to the black hole size (which turns out to be
so small that for the computations that follow we can approximate it as zero)
to a maximum separation $x=\bar x$.  We also assume that the BHMACHOs' motions,
if any, relative to the primordial gas have been redshifted to neglible speeds
by the time their mutual gravitational attractions become important.

Consider a pair of black holes with mass $M_{BH}$ and
a comoving separation $x < \bar{x}$. 
These holes' masses produce a mean energy density over a sphere the size of
their separation given by
$
\rho_{BH}\equiv \rho_{eq}(\bar x/x R)^3. 
$  
This becomes larger than the radiation energy density
$
\rho_r ={\rho_{eq}}/{R^4}
$ 
for 
\be
R>R_m\equiv \left({x/\bar x}\right)^3.
\ee
This means that the binary decouples from the cosmic
expansion and becomes a bound system when $R=R_m$. 
Note that the background universe is still 
radiation dominated at this stage. 

If the motion of the two black holes is not disturbed, 
then the binary system cannot 
obtain any angular momentum so 
they coalesce to a single black hole on the free fall
time scale. 
However, the tidal force from 
neighboring black holes gives the binary enough
angular momentum to keep the holes from colliding with each other 
unless $x$ is exceptionally small. 

We refer to the semimajor axis and the semiminor axis 
of the binary as $a$ and $b$, respectively, and we estimate 
$a$ as
\be 
a = x R_m={x^4/ \bar x^3},  
\ee
We denote by $y$ the comoving separation of the nearest neighboring 
black hole from the center of mass of the binary. 
Then $b$ can be estimated as (tidal force)$\times$(free fall 
time)$^2$ 
\be
b= \frac{GM_{BH}\,xR_m}{(y R_m)^3} 
      \,{(xR_m)^3\over GM_{BH}}
    = \left({x\over y}\right)^3 a.
\ee
Hence, the binary's eccentricity $e$  is given by 
\be
 e = \sqrt{1- \left({x/ y}\right)^6}.
\ee

Since (by assumption) $x$ and $y$ have uniform 
probability distributions in the range $x<y<\bar x$,
the probability distribution of $a$ and $e$ 
is 
\ba
f(a,e)\, da\, de &=& 18x^2y^2 \bar{x}^{-6}\, dx\, dy, \cr
 &=& 
 (3/2)a^{1\over 2}{\bar x^{-{3\over 2}}
 e(1-e^2)^{-{3\over 2}}}\, da\, de. 
\label{fae}
\ea
{}From the condition that $y<\bar x$, 
the maximum value of the eccentricity for a fixed $a$ is given by 
$e_{max}=\sqrt{1-(a/\bar x)^{3\over 2}}$. Integrating $f(a,e)$ with
respect to $e$, we obtain the following distribution of the semimajor axis 
\be
f_a(a)\, da={3\over 2}\left[\left(\frac{a}{\bar x}\right)^{3\over 4}
         -\left(\frac{a}{\bar x}\right)^{3\over 2}\right]
\frac{da}{a}
\label{fa}
\ee

{}From Eq.~(\ref{fa}),  
it is found that the fraction of BHMACHOs that are in binaries with  $a\sim
2\times 10^{14}$cm is $\sim 8\%$ and $\sim 0.9\%$ for
$\Omega h^2=$ 1 and 0.1, respectively.  This estimated fraction of
$\sim $10 AU size BHMACHO binaries is 
slightly smaller than the observed rate of binary MACHO events 
(one binary event in 8 observed MACHOs), but the agreement is good given the
statistics of small numbers.
 
\section{ Gravitational Waves from Coalescing BHMACHO Binaries}
We consider here short period BHMACHO binaries. 
Their coalescence times due to the emission of gravitational waves are
approximately given by (\cite{peter63,peter64}) 
\be
 t=t_{0}\left({a\over a_{0}}\right)^4(1-e^2)^{7\over 2}, 
\label{GWt}
\ee
where $t_{0}=10^{10}{\rm year}$ and 
\be
 a_{0}=2\times 10^{11}\left({M_{BH}\over \sol}\right)^{3\over 4}\hbox{\rm cm}
\label{defa0}
\ee
is the semimajor axis of a binary with circular orbit which 
coalesces in $t_{0}$.  
Integrating Eq.~(\ref{fae}) for a fixed $t$ 
with the aid of Eq.~(\ref{GWt}), 
we obtain the probability distribution for the coalescence time  
$f_t(t)$ as 
\begin{equation}
 f_t(t)dt = {3\over 29}
  \left[\left({t\over t_{max}}\right)^{3\over 37}-
       \left({t\over t_{max}}\right)^{3\over 8} \right]{dt\over t},
\end{equation}
where 
$
 t_{max}=t_0\left({\bar x/ a_0}\right)^4. 
$ 
If the halo of our galaxy consists of BHMACHOs of mass $\sim
0.5\sol$, $\sim 10^{12}$ BHMACHOs exist out to the LMC. The number of
coalescing binary BHMACHOs with $t\sim t_0$ then becomes  
$\sim 5\times 10^8$  for $\Omega h^2 =0.1$ so that the event 
rate of coalescing binaries becomes $\sim 5\times 10^{-2}$events/year/galaxy.
If, however, the BHMACHOs extend up to half way to M31, the number of 
coalescing binary BHMACHOs with $t\sim t_0$ 
can be $\sim 3\times 10^9$ and the event rate becomes 
 $\sim 0.3$ events/year/galaxy.  Both of these estimates are much larger than
the best estimate of the event rate of coalescing neutron stars
based on the statistics of binary pulsar searches in our galaxy, $\sim 1\times
10^{-5}$events/year/galaxy
(\cite{phin91,nara91,vdhe96}). 

The detectability of these waves by interferometers is most easily discussed in
terms of the waves' ``characteristic amplitude'' $h_c$ (Eq.\ (46b) of
Thorne 1987, 
with a well-known factor 2 correction): 
\begin{equation}
h_c=4\times 10^{-21}\left(\frac{M_{chirp}}{M_\odot}\right)^{5/6}
\left(\frac{\nu}{100Hz}\right)^{-1/6}\left(\frac{r}{20Mpc}\right)^{-1} \;.
\end{equation}
Here $M_{chirp} = (M_1 M_2)^{3/5} / (M_1 + M_2)^{1/5}$ is the ``chirp mass'' of
the binary whose components have individual masses $M_1$ and $M_2$.  This $h_c$
is to be compared with an interferometer's ``sensitivity to bursts'' 
$h_{SB} = 11 [f S_h(f)]^{1/2}$, where $S_h(f)$ is the spectral density of the
interferometer's strain noise (cf.\ Eq.\ (111) of Thorne 1987, where 
$h_{SB}$ is denoted $h_{3/yr}$).
  This $h_{SB}(f) \equiv h_{3/yr}(f)$ is plotted
in various publications, e.g. Abramovici et. al. (1989) and Thorne (1995).  It
has a minimum (optimal sensitivity) at a frequency $f \simeq 100$ Hz.  For the
first LIGO and VIRGO interferometers, which are expected to be operational in
2001, that minimum is $h_{SBmin} \simeq 3\times 10^{-21}$.  The GEO600 and TAMA
interferometers, with their somewhat shorter armlengths, will have
$h_{SBmin}$ a little worse than this in 2001.  
LIGO/VIRGO should be able to detect coalescing
binaries, with high confidence, out to the distance for which $h_c = h_{SBmin}$
at the optimal frequency $f \simeq 100$ Hz.  Inserting $M_1 = M_2 \simeq 0.5
M_\odot$ for BHMACHOs into the above equations, we see that {\sl the first
LIGO/VIRGO interferometers in 2001 should be able to see BHMACHO coalescences 
out to about 15Mpc distance}, i.e., out to the VIRGO cluster, 
where our estimates ($\sim$ 1/100 years in each galaxy like our own) 
suggest an event rate of several per year.

LIGO R\&D for the first interferometers is now nearing completion and is
beginning to be redirected toward interferometer enhancements,  
for which the sensitivity goal is a factor 10 improvement, to $h_{SBmin} 
\simeq 3 \times 10^{-22}$ (Barish et.\ al.\ 1996).  In the mid
2000s, with these enhancements in place, LIGO should be able to see BHMACHO
coalescences out to about 150Mpc, which would give a few events per year even
if the event rate is 1000 times smaller than our estimates, 
$\sim 10^{-5}$events/year/galaxy.

\section{Discussion} 
 In this {\it Letter} we have estimated the distribution function of 
binary BHMACHOs in order of magnitude. 
It is possible to compute the distribution function more accurately by
N-body numerical simulations. 
This is an important, challenging numerical problem. 

Our estimated event rate for coalescing BHMACHO binaries 
is comparable to or greater than the most optimistic upper limit for binary 
neutron star coalescences (\cite{phin91}), which are one of the most important 
sources of gravitational waves. Coalescing neutron stars
are also regarded as possible sources of the gamma-ray
bursts (\cite{mesz95}). If so, then the detection of
gravitational waves should be accompanied by a gamma-ray burst. 
If we consider the fire ball model (\cite{mesz95})
the time delay between  the gravitational waves and 
the gamma-rays should be $\sim$ 1 sec. 
By contrast, in the coalescence of binary BHMACHOs the
emission of gamma-rays is not expected.
This may enable us to distinguish
coalescing binary BHMACHOs from coalescing binary neutron stars.

If gamma-rays are not emitted by coalescing binary neutron stars, we
may still use their observed chirp masses $M_{chirp}$
to distinguish them from BHMACHO binaries. The chirp mass can be measured from
the gravitational waves to a fraction of a per cent accuracy, which is
much less than the expected spreads of BHMACHO masses and neutron star masses. 
The masses of neutron stars in
binaires are expected to be $\sim 1.4\sol$ corresponding to a chirp mass of
$1.2\sol$.  This is supported observationally as well as theoretically 
since the mass of the
iron core before the collapse is $\sim 1.4\sol$. 

Therefore if the chirp masses
of BHMACHO binaries are much smaller than $1\sol$ it is possible to 
identify them.
However if the mass of a BHMACHO is $\sim 1\sol$, 
coalescing BHMACHO binaries and coalescing neutron star binaries
may be indistinguishable from the detected gravitational waves. 
In principle, however, black holes absorb some of the
gravitational waves so that the evolution of the binary
 in the last few minutes
(\cite{cutl93}) is slightly different from that of a neutron star binary 
of the same mass. The difference arises in the 2.5th post-Newtonian order
for the Kerr black hole case (\cite{tago97}) and 4th order for the
Schwarzschild black hole case (\cite{PoSa95}). This difference might be
detectable, though it will present a
challenging problem for gravitational wave data analysis.   
  
It is known that the Silk damping of density perturbations on 
small scales causes distortions of the CMB spectrum by dumping 
acoustic energy into heat and thence into the CMB (\cite{Daly91,Barr91,Hu94}).
Thus it is important to examine if the large
primordial density perturbations that are needed for 
BHMACHO formation are compatible with the observed upper limit of
CMB spectral distortions (\cite{Math94}).

For a rms amplitude of density perturbations $\delta(l)$ on a comoving
scale $l$ in the radiation dominated era,
the fraction of the universe that turns into black holes is
given by $f(l)\approx\delta(l)\exp\left(-{1/18\delta(l)^2}\right)$ 
(\cite{Carr75}). 
In the present case, $f(l)$
is equal to $R_f$ given by Eq.~(\ref{Rf}) and $l$ is equal to
the comoving scale $l_{BH}$ of the density perturbations
 that give rise to BHMACHOs. 
Recalling that we fixed the scale factor $R$ to unity at 
the time $t_{eq}$ of matter-radiation equality, 
the present comoving scale corresponding to $l_{BH}$ is given by 
$(1+z_{eq})l_{BH} =
(1+z_{eq}) R_f^{-1}GM_{BH}c^{-2} = 
5\times10^{17}\left({M_{BH}/\sol}\right)^{1/2}{\rm cm}$, 
where $z_{eq}$ is the redshift at $t=t_{eq}$. 
Since $R_f\sim 10^{-8}$, we need $\delta(l_{BH})^2\sim 4\times 10^{-3}$.

On the other hand, Hu, Scott and Silk (1994) showed that the
observational limit constrains the
rms amplitude of density perturbations at horizon entry to be
$\delta(l)^2\lesssim10^{-4}$ for $l>l_D$, 
where $l_D$ is the comoving scale of the Silk damping 
at the double Compton thermalization time,
which at present is given by
$(1+z_{eq})l_D=3\times10^{20} {\rm cm}$. 
Hence the scale of interest $l_{BH}$ is approximately 3 orders of magnitude
smaller than the scale $l_D$ to which the observational constraint applies.

Correspondingly, if we assume the primordial density perturbation spectrum
to have a power-law shape, $P(k)\propto k^n$, on scales greater 
than $l_{BH}$, the above result implies that $n\gtrsim1.6$. 
Interestingly, this value is consistent with the one
suggested by the observed CMB anisotropies (\cite{Benn94}) 
and it is marginally
consistent with the COBE normalization (\cite{Hu94}). 
It may be that this rather blue spectrum can be produced by a variant of the 
so-called hybrid model of inflation(\cite{Garc96}).

\acknowledgments
 This work was supported in part by a Japan-US cooperative program on
gravitational waves from coalescing binary compact objects (JSPS grant
EPAR/138 \&
NSF grant INT-9417348). The main
part of the paper was completed while the Japanese authors 
(TN, MS and TT) were visiting Caltech. 
 This work was also supported by a
Grant-in-Aid for Basic Research of the Ministry of Education,
Culture, and Sports No.08NP0801,09640351 and by NSF grant AST-9417371.

\end{document}